\documentclass[aps,prl,floatfix,singlecolumn,showpacs]{revtex4}
\usepackage{graphicx} 
\usepackage[sort&compress]{natbib}

\newcommand{\B}{{\rm B}}
\newcommand{\BH}{{\rm BH}}
\newcommand{\BEQ}{\begin{equation}}
\newcommand{\EEQ}{\end{equation}}
\newcommand{\BEA}{\begin{eqnarray}}
\newcommand{\EEA}{\end{eqnarray}}
\begin{document}

\title {  Role of Various Entropies in the
 Black Hole \\
 Information Loss Problem
}

\author{Th.M. Nieuwenhuizen$^{1)}$ and I.V. Volovich$^{1,2)}$ $~~~~$\\
$~~~~~~~~~$\\
Institute for Theoretical Physics$^{1)}$,\\
Valckenierstraat 65, 1018 XE
Amsterdam, The Netherlands\\
$~~$\\
Steklov Mathematical Institute$^{2)}$,\\
 Gubkin St. 8, 119991, Moscow,
Russia\\
email: volovich@mi.ras.ru}



\begin{abstract}
In a recent paper Hawking has argued that there is no information
loss in black holes in asymptotically AdS spacetimes. We remind
that  there are several types of information (entropy) in
statistical physics -- fine grained (microscopic) and coarse
grained (macroscopic) ones which behave differently under unitary
evolution. We suggest that the coarse grained information  of the
rest of the Universe is lost while fine grained information is
 preserved.  A
possibility to develop in quantum gravity an analogue of the
Bogoliubov derivation of the irreversible Boltzmann and Navier -
Stokes equations from the reversible mechanical equations is
discussed.
\end{abstract}

\pacs{04.70.Dy}

\maketitle



\section{Introduction}

In 1976 Hawking has  argued that the black hole creation and
evaporation could lead to an evolution of a pure state into a mixed
state, which is in  contradiction with the rules of quantum
mechanics ~\cite{Haw76}. This has become known as the black hole
information loss problem (or black hole information paradox), for a
review see \cite{FroNov}.

In a recent paper Hawking ~\cite{Haw05} has suggested that there is
no information loss in black holes in asymptotically AdS (anti-de Sitter)
spacetimes. The central point in his argument is the assertion
that there is no information loss if one has a unitary evolution.
Then, since the AdS quantum gravity is dual to a unitary conformal
field theory ~\cite{Mal,GKP,Wit}, there should be no information loss.

In more detail the proposal looks as follows.  Black hole
formation and evaporation is considered as a scattering process
when all measurements are made at infinity. The evolution operator
is defined by means of the path integral for the partition
function. In the sum over topologies in the path integral there
are trivial topologies, which lead to the unitary evolution, and
nontrivial black hole topologies. Information is lost in
topologically non-trivial black hole metrics, but their
contribution decays to zero at large times, at least for each
separate mode,\cite{Mal2}. As a result one gets a unitary
evolution at large times and the information is preserved in that
limit.

 We make two remarks concerning this proposal.

  (i) Actually it is proposed that we can make observations only at
such large time scales where small black holes either do not form
yet or have already evaporated. In any case black holes are not present then.

 (ii) An important point in the argument is the assertion that, if one
considers a theory with a unitary dynamics, there is no
information loss problem.

Concerning the point (i)  there is a question of how information
gets out of a black hole. In fact it is one of the main questions in
the whole discussion of the information paradox. Hawking's answer is
that  there is no sense in asking this question because its answer
would require to use a semi-classical metric, which has already lost
the information. We discuss this argument below.

About the point (ii) we remind  that  the problem of how a unitary
reversible dynamics can lead to an irreversible behavior (i.e. to
information loss) is the subject of   numerous studies in
statistical physics. One specific application is the problem of
relaxation: Even though quantum evolution, believed to describe all
condensed matter, is unitary, we observe relaxation in every day
life~\cite{Balian}. It is thus clear that much of the irreversible
behavior, including probably the expansion of the Universe, should
be compatible with unitary quantum mechanics. Point (ii) is the main
point which will be discussed in this note.

We consider the black hole information problem as a particular
example of the fundamental irreversibility problem in statistical
physics. We point out that similar problem occurs when we study
ordinary gas or the formation of the ordinary black {\it body} and
its thermal radiation. Actually, one has to give a quantum
mechanical explanation for the emergence of the second law of thermodynamics
in macroscopic systems.

The irreversibility problem was much studied by Boltzmann and many
other authors. There is not yet a complete solution  but a deep
understanding of the problem has been achieved \cite{Bog} -
\cite{Gal}.

Information is usually quantified as entropy with a minus sign.
There are two classes of entropies in statistical physics -- the
fine grained (microscopic) one and several coarse grained
(macroscopic) ones. They behave differently under unitary evolution.
For a given model, unitary dynamics might increase one or more
coarse grained entropies while preserving the fine grained one.
Whether there is an increase of a certain coarse grained entropy
(i.e. loss of {\it this} information) is a dynamical question and
its answer comes for a given model from a thorough investigation of
its dynamics, which should show a sort of instability or ergodicity
and mixing. Alternatively, coupling to a bath would suffice. This is
relevant for the black hole situation, where the ``bath" is the rest
of the Universe.

We note that the properties of the coarse grained (in particular
thermodynamical) entropy are important in statistical physics while
the fine grained entropy is not so significant since typically it
cannot be determined  and, remaining conserved, would not reflect a
specification of the dynamical evolution.

\section{Different kinds of entropies}

Let us split up our degrees of freedom in two classes, those of the
black hole and those of the rest of the Universe. The latter we call
``bath'', since for a proper formulation of thermodynamics of black
holes, the rest of the Universe indeed plays the role of the thermal
bath in condensed matter problems. The density matrix of the total
system $\rho$ has several marginals. The reduced density matrix of
the black hole is $$ \rho_\BH={\rm Tr}_\B\,\rho $$ while the reduced
density matrix of the bath is $$\rho_\B={\rm Tr}_\BH\,\rho$$ This
brings us three von Neumann entropies: the ``fine grained'' entropy
of the total system, $$ S_{\rm total}^{{\rm fine}}=-{\rm
Tr}\,\rho\,\ln\rho. $$ This quantity is conserved in time for
unitary motion. If one starts out from a pure state of incoming
matter and bath, it vanishes at all times. Next there is the fine
grained von Neumann entropy of the black hole, $$ S_{{\BH}}^{{\rm
fine}}=-{\rm Tr}_\BH\,\rho_\BH\,\ln\rho_\BH
 $$
and the fine grained von Neumann entropy of the bath, $$
S_{{\B}}^{{\rm fine}}=-{\rm Tr}_\B\,\rho_\B\,\ln\rho_\B
 $$

When starting from a pure state, entangled or not,  the latter two
will be equal at all times. Though then vanishing at $t=0$ they
become positive at later times, $S_{{\BH}}^{{\rm
fine}}(t)=S_{{\B}}^{{\rm fine}}(t)>0$. At large times one expects
them to go to zero again~\cite{however}. For  $S_{{\BH}}^{{\rm fine}}$ the reason
is simply that matter is radiated, making it smaller and smaller,
 so that with the matter its entropy evaporates.
For $S_\B^{{\rm fine}}$ it theoretically expected, because, in the
final absence of the hole, it just reflects the purity of the
state. But physically it is a surprising and counter-intuitive
result, since this vanishing entropy clearly does not reflect the energy
radiated into this bath by the black hole evaporation.

Coarse graining can be done at larger and larger scales. Thus
there is still another entropy to consider, namely the coarse
grained entropy of the bath, where one neglects all correlations
between bath modes. One first has to define the reduced density
matrix of a given mode, $\rho_{\rm mode}={\rm Tr}_{\rm
all\,\,other\,\,modes}\,\,\rho$ and from it the von Neumann
entropy $$ S_\B^{\rm coarse}=\sum_{\rm mode} S_{\rm mode}=
\sum_{\rm mode}-{\rm Tr}_{\rm mode}\,\,\rho_{\rm mode}\ln\rho_{\rm
mode} $$ This is the one entering quasi-classical discussions of
the black hole information paradox. Since it probably does not
vanish at large times, it is a candidate for our ``natural''
association of entropy with a measure of disorder. It is then
obvious that $ S_{{\B}}^{{\rm fine}}$, which does vanish at large
times, takes the correlations between different modes, neglected
in $S_\B^{\rm coarse}$, into account in a very subtle manner.

Boltzmann's  famous formula for the entropy reads
$$
S=k_B\log W
$$
The W in this formula is the number of microstates compatible with
the {\it macroscopic} state. Therefore this formula defines a coarse
grained entropy. There is a remarkable computation by Strominger and
Vafa of the Bekenstein - Hawking coarse grained  entropy by counting
of microscopic BPS states in string theory \cite{SV}.

Black hole thermodynamics is a problem with two temperatures: the
Hawking temperature of the hole and the $3K$ back ground temperature
~\cite{TNBH}~\cite{TN3rdlaw}. These distinctions automatically show
up in condensed matter analogs of the black hole evaporation
problem, that we plan to discuss elsewhere~\cite{NVA}.

The thermodynamic entropy of a system in contact with a bath at
temperature $T$ is defined in terms of added heat $d Q$ by the
Clausius inequality \BEQ d Q\le T d S \EEQ
 taken as an
equality, so $d S=d Q/T$. The ``classical intuition" of entropy
arose when Bolztmann showed that this thermodynamic entropy agrees
with his measure of disorder, more precisely, the logarithm of the
number of states. For a closed system, this quantity cannot
decrease. For systems with two temperatures a generalized Clausius
inequality may hold when there are also two well separated time
scales, leading to two different entropies, that enter as: $d Q\le
T_1 d S_1+T_2 d S_2$. This applies to glasses~\cite{TNglass} and
black holes~\cite{TNBH}.

For nanoscopic and mesoscopic systems, one has led to uncover the
field of ``quantum thermodynamics"~\cite{qThermo}.

Page has discussed the microcanonical and canonical entropies of
black holes~\cite{Page04}. The first one does not reflect its
environment, the second one assumes it to be at the Hawking
temperature, which is the case only for a specific black hole size
and, moreover, an unstable situation.

We have discussed various entropies which are used to describe the
{\it classical} capacity of a quantum channel. To describe the {\it
quantum} channel capacity the  quantum mutual entropy \cite{Ohy}
and the quantum
 coherent information \cite{NC} are used. It would be interesting to
investigate the role of these entropies in quantum gravity.

\section{On  the black hole information paradox}

 Our proposal for the investigation of the black hole
information loss problem is the following. One of the mentioned entropies is
the  coarse grained bath entropy.  It increases during the
evolution. In this sense there is information loss. But it does
not mean that there is loss of the fine grained information
(entropy). The whole picture of the black hole formation and
evaporation is similar to the formation and radiation of a black
body.

In a specific model the increase of coarse grained entropy has to
be demonstrated if the black hole evaporation indeed behaves as a
thermodynamic problem.

The next step in the program is to develop in quantum gravity an
analogue of the Bogoliubov derivation of the Boltzmann and Navier
- Stokes equations from the Liouville equation~\cite{Bog,Isi,ZMR}.
We shall sketch  that in the last section. In quantum field theory
one derives quantum stochastic differential equations \cite{ALV}.
But we should stress that a much better understanding of the
irreversibility problem in various models is required.

It would also be interesting to investigate the role of the
quantum mutual entropy \cite{Ohy} and of
the quantum coherent information \cite{NC} in quantum gravity.

\section{Information Loss in Gases}

Consider a classical or quantum gas of $N$ particles in a box of the volume $V$ which
is described by  the Hamiltonian $$
 H_N=\sum_{i=1}^{N}\frac{{\bf
p}_i^2}{2m} + \sum_{i<j} \Phi({\bf x}_i - {\bf x}_j). $$
 Here
${\bf x}_i$ are positions, ${\bf p}_i$ are momenta,  $m$ is mass
and $\Phi ({\bf x})$ is the interaction potential between a pair of
particles.

One has a reversible classical dynamics and a unitary quantum
dynamics. Hence, in neither situation there is a loss of the fine grained information.

However, it is well known that the kinetic theory of gases is
based on the Boltzmann equation which reads \cite{Isi,ZMR}
\begin {equation}
\label{B} \frac{\partial f}{\partial t}+\frac{{\bf
p}}{m}\cdot\frac{\partial f}{\partial {\bf x}}+{\bf F}\cdot
\frac{\partial f}{\partial {\bf p}}=J(f).
\end{equation}
 Here $f=f({\bf x},{\bf p},t)$ is the one-particle distribution
function,  $t$ is time, ${\bf F}={\bf F}({\bf x},t)$ is the force,
and $J(f)$ is a bilinear functional in $f$.

The coarse grained Boltzmann entropy is defined by $$ S_B(t)=-\int
f({\bf x},{\bf p},t)\ln f({\bf x},{\bf p},t) d{\bf x} d{\bf p}.
$$
Boltzmann has proven ($H$-theorem) that
$$ \frac{d S_B(t)}{d t}\geq 0 $$
 Moreover the entropy is constant
only for an equilibrium distribution function. For a
non-equilibrium state the Boltzmann entropy increases and
therefore one gets {\it information loss}.

Now one can ask the same question as for the black hole
information loss problem. How is it possible that one gets
information loss and irreversibility for a system of $N$
particles which is described by a reversible dynamics?

An important progress in investigation of this question was
achieved by Bogoliubov \cite{Bog,Isi,ZMR}. He considers a system
of equations for $s$-particle correlation functions
$$
f_s(\xi_1,...,\xi_s,t)
$$
 where $\xi_i=({\bf x}_i,{\bf p}_i),$ ~
$i,s=1,2,...,N$. The system of equations (BBGKI-chain) is
equivalent to the Liouville equation and is reversible.

Then for the dilute gases Bogoliubov introduces a kinetic
relaxation time scale $\tau_0$ and uses the thermodynamical limit
($N, V\to\infty,~ N/V=$ const) and the factorization of the $s$-
particle correlation functions $f_s$ in terms of the one particle
distribution function $f$:
\begin {equation}
\label{1} f_s(\xi_1,...,\xi_s,t)\to\prod_{i=1}^s
f(\xi_i,t),~~~t>\tau_0,~~s=2,3,...,N
\end{equation}

 In this way he was able to obtain the Boltzmann
equation. Then one can use the Boltzmann equation to derive the
hydrodynamical Navier - Stokes equation.

We write the Boltzmann equation (\ref{B}) symbolically as
\begin{equation}
\label{lB} \frac{\delta f}{\delta\sigma}=J(f)
\end{equation}
where $\sigma =(t,\xi)$.

Bogoliubov used a similar approach also for derivation of quantum
kinetic equations. In this case one uses the correlation functions
$$ f_{sk}({\bf x}_1,...,{\bf x}_s;{\bf y}_1,...,{\bf y}_{k};t)
=Tr[\rho_t \psi ({\bf x}_1)...\psi ({\bf x}_s)\psi^+ ({\bf
y}_{1})...\psi^+ ({\bf y}_k)] $$ Here $\rho_t$ is the density
operator at time $t$ and $\psi ({\bf x}),~\psi^+ ({\bf y})$ are
annihilation and creation operators satisfying the usual
commutation relations $ [\psi ({\bf x}),\psi^+ ({\bf y})]=\delta
({\bf x} - {\bf y}). $ To derive quantum kinetic equation one uses
an approximation similar to the Bogoliubov approximation
(\ref{1}). It is a kind of the mean field approximation.

 There are other models in which an
irreversible behavior from reversible dynamics was derived,
\cite{ALV, Leb}. Mathematical studying of these questions is also
the subject of ergodic theory where for certain dynamical systems
the properties of ergodicity and mixing were established and where
the notions of classical and quantum Anosov and K-systems and the
Kolmogorov - Sinai entropy play an important role
\cite{Ano,Sin,AKN,Gal}.

It was demonstrated by Pauli,  von Neumann, van Hove, and Prigogine
that in quantum mechanics the coarse grained entropy increases as
a result of the unitary dynamics.

 The transition from the BBGKI chain of equations for the
family of correlation functions $\{f_s\}$ to the Boltzmann
equation for the one particle distribution function $f$ is the
transition from the fine grained reversible description to the
irreversible coarse grained description. We loose information when
we describe a gas by means of only the one-particle distribution function
$f({\bf x},{\bf p},t)$, or hydrodynamical variables, which are
integrals of $f$ with some weights.

In quantum gravity we interpret the distribution $f[g,\pi]$ of the
classical metric $g_{\mu\nu}(x)$ and its conjugate
$\pi_{\mu\nu}(x)$ as an analogue of the one particle distribution
$f({\bf x},{\bf p},t)$ or hydrodynamical variables. To get insight
into the  black hole information loss problem, one has to develop
in quantum gravity an analogue of the Bogoliubov derivation of the
Boltzmann equation from the Liouville equation. In quantum field
theory one derives quantum stochastic differential equations
\cite{ALV}.

\section{Information Loss in Quantum Gravity}
One can use the Euclidean \cite{Haw05} or  the Wheeler - De Witt
\cite{BG} approach to quantum gravity. Each one has its own
advantages and disadvantages. One of problems with the Euclidean
approach is that one can define there the Green functions and the
partition function but not the scattering matrix. Moreover, path
integrals are convenient to write down semiclassical expansion but
one can not use this "sane" formalism to solve  spectral problems,
even to compute the spectrum of the hydrogen atom.

The transition amplitude between   configurations of the
three-metric $h_{ij}'$ and field $\Phi '$ on an initial spacelike
surface $\Sigma ' $ and a configuration $h_{ij}''$ and  $\Phi ''$
on a final surface
 $\Sigma '' $ is
 $$
<h'',\phi '', \Sigma ''|h', \phi ' , \Sigma '>= \int ~~
e^{\frac{i}{\hbar}S[g,\Phi]} {}~{\cal D}\Phi {\cal D}g,
$$
where the integral is over all four-geometries and field
configurations which match  given values on two spacelike
surfaces, i.e. $\Phi |_{\Sigma '}=\phi ',~ g |_{\Sigma '}=h ' ,$
$\Phi |_{\Sigma ''}=\phi '',~ g |_{\Sigma ''}=h ''$.

The problem of creation of black holes in quantum theory is
considered in \cite{tHo} - \cite{Rin}. The role of boundary
conditions in the path integral describing the creation of black
holes is discussed in \cite{ArefVisVol}.

We are interested in the process of black hole creation. Therefore
$\Sigma '$  is a partial Cauchy surface with
 asymptotically simple
past in a strongly asymptotically predictable space-time and
$\Sigma ''$  is a partial Cauchy surface
 containing black hole(s), i.e.
$\Sigma ''-J^{-}({\cal T}^{+})$ is non empty.

Black holes  are conventionally defined \cite {HE} in
asymptotically flat (or AdS) space-times by the existence of an
event horizon $H$. The horizon $H$ is the boundary $\dot
{J}^{-}({\cal I}^{+})$
 of the causal past $J^{-}({\cal I}^{+})$
of future null infinity ${\cal I}^{+}$. The black hole region  $B$
is $B=M-J^{-}({\cal I}^{+})$ and  the event horizon $H=\dot
{J}^{-}({\cal T}^{+}).$ This definition depends on the whole future
behavior of the metric. There is  a different sort of horizon,
trapped horizon, which depends only on the properties of space-time
on the surface $\Sigma (\tau) $ \cite {FroNov,HE}.

We discussed the transition amplitude (propagator) between
definite configurations of fields, $<h'',\phi '', \Sigma ''|h',
\phi ' , \Sigma '>$. The transition amplitude from  a state
described by the wavefunction $\Psi ^{in}[h', \phi ']$ to a state
$\Psi ^{out}[h'', \phi '']$ reads
$$
<\Psi ^{out}|\Psi ^{in}>=
$$
$$ \int \bar{\Psi }^{out}[h'', \phi ''] <h'',\phi '', \Sigma ''|h'
,\phi ' , \Sigma '>\Psi ^{in}[h' ,\phi '] {\cal D}h'{\cal D}\phi
'{\cal D}h''{\cal D}\phi ''.$$

Consider a family of asymptotically flat spacetimes. A wave
function is a functional of the 3-geometry and the matter fields
$\Psi [h_{ij},\phi]$. It satisfies the Wheeler - De Witt equation
$$ {\cal H}\Psi =0 $$ where ${\cal H}$ is the density of the
Hamiltonian constraint.

Let $\rho_{\Sigma}$ be the density operator of the Universe at the
surface ${\Sigma}$. One defines the fine grained entropy
$$
S^{{\rm fine}}(\Sigma)=-{\rm Tr}\,\rho_{\Sigma}\,\ln\rho_{\Sigma}
$$
and  correlation functions
$$
f^{(s)}_{i_{1}j_{1}...i_{s}j_{s}}(x_1,...,y_s;\Sigma)=
Tr[\rho_{\Sigma} {\hat g}_{i_{1}j_{1}}(x_1)...{\hat
\pi}_{i_{s}j_{s}}(y_s)]
$$
where ${\hat h}_{ij}, {\hat \pi}_{ij}$  operators of metric and
its canonically conjugate.

The correlation functions satisfy a system of equations in
superspace. The equations  are complicated. We can assume that
there is a sort of unitary dynamics which preserves the fine
grained entropy but there is no way to determine it. This is
similar to the gas dynamics which was discussed in the previous
section. But let us try to derive an analogue of the Boltzmann
equation. We consider an approximation:
$$
f^{(s)}_{i_{1}j_{1}...i_{s}j_{s}}(x_1,...,y_s;\Sigma)\to\prod_r
f_{i_{r}j_{r}m_{r}n_{r}}(x_r,y_r;\Sigma),~~\Sigma
>\Sigma_0,
$$
where $$ f_{i_{r}j_{r}m_{r}n_{r}}(x_r,y_r;\Sigma)
 =Tr[\rho_{\Sigma} {\hat
g}_{i_{r}j_{r}}(x_r){\hat \pi}_{m_{r}n_{r}}(y_r)]
$$
This is similar to the Bogoliubov approximation (\ref{1}). Quantum
Boltzmann equation in quantum gravity will have the form
$$
 \frac{\delta f}{\delta\sigma}=J(f)$$
where $f=f_{i_{r}j_{r}m_{r}n_{r}}(x_r,y_r;\Sigma)$ and $\sigma
=(x,y,g,\pi,\Sigma)$. The problem is to determine an explicit form
of the functional $J(f)$.

The coarse grained entropy is
$$
 S^{{\rm coarse}}({\Sigma})=-\int tr f(\cdot ; {\Sigma})\ln f(\cdot ;{\Sigma})
$$
where an appropriate normalization for $f$ is assumed.

 Further, if we make also the
semiclassical approximation, or assume the coherent pure states,
then, in principle,  we could get the classical Einstein equations
for the metric $g_{\mu\nu}(x).$ However the form of classical
equations depend on the chosen state.

For classical gravity the known laws of black hole thermodynamics
are valid. In this case the entropy, which  is proportional to the
area of horizon, is a coarse grained entropy. This entropy
increases during the  classical evolution, so one gets a loss of the
coarse grained information.

We speculate that the coarse grained entropy which is obtained in
the Bogoliubov approximation  increases  in the classical regime.
Note that the Hamiltonian in mini-superspace \cite{BG} is an $N$-
particle hamiltonian but it has a form more complicated then the
Hamiltonian for gases. One can try to study this problem for the
Matrix model \cite{BFSL}. Note however that the irreversibility
problem is a rather difficult problem even for such a simple and
well studied system as quantum baker`s map \cite{IOV}.

On later times the evaporation is the only relevant but slow process (quantum regime).
Then the black hole looses matter and its coarse grained
entropy decreases. This matter will go to the bath, so its coarse grained entropy
is expected to increase such that the total coarse grained entropy also increases.

\section{Conclusions}

The information paradox of the black hole problem has bothered
scientists since the discovery of the evaporation process. It has
often not been realized, however, that ``the" entropy of a system
does not exist. The thermodynamic entropy of a system in contact
with a bath at temperature $T$ is defined in terms of added heat $d
Q$ as $d S=d Q/T$. The ``classical intuition" of entropy arose when
Bolztmann showed that this thermodynamic entropy agrees with a
measure of disorder, namely the logarithm of the number of relevant
states. For a closed system, this quantity cannot decrease. On the
other hand, it is known in quantum mechanics that the von Neumann
entropy {\it of a closed system} is a constant due to unitary
motion. This is the quantum analog of the classical fine grained
entropy, which is also conserved in time.

A second complicating factor it that the setup for black hole
thermodynamics is one of systems far from equilibrium~\cite{TNBH}.
Thus the evaporation process is a problem of thermodynamics far from
equilibrium, to which Gibbsian thermodynamics does not apply and for
which, in general, few tools are available.

The problem of black hole evaporation is one of the field of {\it
quantum thermodynamics}: the target system is small (the hole), but
the bath is large (the rest of the Universe) and also the work
source is large (here it would stand for the incoming matter, or
work done externally on the hole), see e.g.\cite{qThermo}. In this
field one can imagine a condensed matter analog where, starting from
a ground state, work is from the outside put into a certain degree
of freedom (`` growing of a toy black hole"), that is later taken
out (its disappearance). In this process, not all work can be
recovered due to Thomson's formulation of the second law: cyclic
processes done on an equilibrium system (here: in its ground state)
cannot yield work, and typically will cost work. This work will end
up as phonons running away in the infinite condensed matter bath, in
the very same way as matter and photons evaporated from the black
hole will run in the otherwise empty, infinite Universe. We plan to
discuss this setup in future, considering the separate entropies in
detail~\cite{NVA}. The recent work on a realistic quantum
measurement~\cite{ABNqmeas} shows that measurement problems are
probably disconnected from black holes issues.

There is a little hope that quantum gravity or string theory could
help to get an insight into the fundamental irreversibility problem
until the further progress in the considerations of simple models
will be achieved. However we should remind that Boltzmann has
predicted the cosmological Big Bang just from the consideration of
the irreversibility problem \cite{Leb}. Not only the black hole
problem but also the recent discovery of the cosmological
acceleration and the mystery of dark energy indicates, it seems, to
the necessity of the unified treatment of the basic problems in
cosmology, quantum gravity/string theory, high energy physics,
statistical physics and quantum information theory.

 As to the black hole problem itself,
we have outlined how to derive an equivalent of the Boltzmann
entropy for gravitation. Also this subject deserves further
attention. One outstanding question is to show that, taken
together with the coarse grained bath entropy, it is
non-decreasing.

\section*{Acknowledgments.}
I.V. V. acknowledges hospitality at the University of Amsterdam,
on a grant by Stichting voor Fundamenteel Onderzoek der Materie
(FOM), financially supported by the Nederlandse Organisatie voor
Wetenschappelijk Onderzoek (NWO). He is grateful to  A. Arvilsky
and B. Dragovich for fruitful discussions and to grants RFFI
05-01-00884 and NSCH-1542.2003.1 for support.


\begin{thebibliography}{72}
\bibitem{however} There are also scenario's in ~\cite{NVA} where the
toy-black-hole entropy remains constant, or even diverge logarithmically
with time.
\end{thebibliography}
\end{document}